\documentclass[checkin,aps,pra]{revtex4}
\usepackage{makeidx}
\usepackage{listings}
\newcommand{\ba}{\begin{eqnarray}}
\newcommand{\ea}{\end{eqnarray}}

\usepackage{graphicx,color}

\newcommand{\be}{\begin{equation}}
\newcommand{\ee}{\end{equation}}
\newcommand{\la}{\langle}

\newcommand{\ra}{\rangle}
\newcommand{\et}{{\it et al. }}

\newcommand{\clr}{}

\begin{document}




\title{Spin-phonon dispersion in magnetic materials}





\author{Mingqiang Gu} \affiliation{Department of Physics, Southern
  University of Science and Technology, Shenzhen 518055, China}


\author{Y. H. Bai} \affiliation{Office of Information
  Technology, Indiana State University, Terre Haute, Indiana 47809,
  USA}

\author{G. P. Zhang$^*$} \affiliation{Department
  of Physics, Indiana State University, Terre Haute, Indiana 47809,
  USA}

\author{Thomas F. George}

\affiliation{Departments of Chemistry \&
  Biochemistry and Physics \& Astronomy, University of
  Missouri-St. Louis, St.  Louis, MO 63121, USA }

\date{\today}

\begin{abstract}
  {Microscopic coupling between the electron spin and the lattice
    vibration is responsible for an array of exotic properties from
    morphic effects in simple magnets to magnetodielectric coupling in
    multiferroic spinels and hematites.  Traditionally, a single
    spin-phonon coupling constant is used to characterize how
    effectively the lattice can affect the spin, but it is hardly
    enough to capture novel electromagnetic behaviors to the full
    extent. Here, we introduce a concept of spin-phonon dispersion to
    project the spin moment change along the phonon crystal momentum
    direction, so the entire spin change can be mapped out.  Different
    from the phonon dispersion, the spin-phonon dispersion has both
    positive and negative frequency branches {even in the equilibrium
      ground state}, which correspond to the spin enhancement and spin
    reduction, respectively.  {\clr Our study of bcc Fe and hcp Co
      reveals that the spin force matrix, that is, the second-order
      spatial derivative of spin moment, is similar to the vibrational
      force matrix, but its diagonal elements are smaller than the
      off-diagonal ones.  This leads to the distinctive spin-phonon
      dispersion.  The concept of spin-phonon dispersion expands the
      traditional Elliott-Yafet theory in nonmagnetic materials to the
      entire Brillouin zone in magnetic materials, thus opening the
      door to excited states in systems such as CoF$_2$ and NiO, where
      a strong spin-lattice coupling is detected in the THz regime.} }
\end{abstract}

 \maketitle

 \section{Introduction}

Using the lattice to control the spin degree of freedom has garnered a
significant attention worldwide, as a new way to alter spintronics.
But long before the current surging research activities
\cite{kim2010,zahn2021,prb21}, phonon excitation is known to be
essential to demagnetization in a magnetic material by transferring
energy to the spin subsystem, inducing magnetic phase transition. Even
in nonmagnetic crystals, the magnetic field affects photon-phonon
interactions, which is known as morphic effects
\cite{anastassakis1972}.  The discovery of strong magnetodielectric
coupling in Mn$_3$O$_4$ \cite{tackett2007} demonstrated that the
lattice involvement is more than just energy transfer.  Hirai \et
\cite{hirai2013} reported a large atomic displacement at a magnetic
phase transition, with structural change detected
\cite{kemei2014}. 
Extreme modification on the crystal lattice is expected to change the magnetic ordering and has been demonstrated experimentally with multiferroic (La,Ca)MnO$_3$ \cite{science2021}. 
The Raman scattering experiment \cite{byrum2016}
revealed that there is an opposite effect on the lattice from an
external magnetic field, or magnetostructure change.  A real-space
magnetic imaging of multiferroic spinel was also obtained
experimentally \cite{wolin2018}.  When lattice vibrates, it alters the
exchange interaction \cite{ma2008}.  Theoretically, Lizarraga \et
\cite{lizarraga2017} showed the magnetism itself may play a role in
the phase transition from hcp Co to fcc Co.  Kuang \et
\cite{kuang2014} suggested that Co may have magneto-elastic effects.
Verstraete \cite{verstraete2013} showed that the electron-phonon
coupling is also spin-dependent and the spin minority band has a
larger coupling than the majority one.  Lefkidis \et \cite{georgejmmm}
demonstrated for the first time that the phononic effects also appear
on an ultrashort time scale, through the electron-phonon coupling
\cite{essert2011,baral2014}. The effect becomes stronger in the
rare-earth Gd(0001) \cite{melnikov2003,sultan2012}.  Henighan \et
\cite{henighan2016} directly detected the signature of THz coherent
acoustic phonons in Fe.  To characterize the coupling between the
phonon and spin excitation, a common method is to use the
phonon-magnon coupling or magnetoelastic coupling
\cite{weber1968,guerkeiro1971,cheng2008,mikhail2011,berk2019,sjostrand2021,li2021,vaclavkova2021}.
Pattanayak \et \cite{pattanyak2021} recently detected some of the key
signatures of spin-phonon coupling in hematite crystallites through
dielectric and Raman spectroscopy. However, a single spin-phonon
coupling constant is hardly enough to capture the complexity as how
spin and phonon interact with each other.  The lattice vibration has
$3N-6$ degrees of freedom for a system of $N$ atoms, which is likely
to limit the scope of applications of a single spin-phonon coupling
constant, especially true in magnetostructure and magnetodielectric
coupling.

In this paper, we go beyond a single spin-phonon coupling constant and
introduce a concept of the spin-phonon dispersion. We take the hcp Co
as our first example to show that there is a well-defined nearly
linear dependence of the spin moment on the volume of the cell for a
fixed $c/a$ ratio, and on the $c/a$ ratio for a fixed volume. The
second order in the spin moment change is dispersed along the phonon
crystal momentum. A new picture emerges. Different from the phonon
counterpart, the eigenvalues of the spin-phonon dispersion can be
either positive or negative, which corresponds to the spin moment
increase or decrease, respectively. The orders of the spin-phonon
bands are reversed with respect to the phonon dispersion; and at the
$\Gamma$ point the single-degenerate band appears at the bottom and
the triple-degenerate band at the top. Along the high symmetry lines
in the Brillouin zone, both spin reduction and enhancement are
found. The situation in bcc Fe is different. The major reduction is
around the N point, and there is only spin enhancement along the
$\Gamma$-H line.  We attribute the main differences between the
spin-phonon dispersion and the regular phonon dispersion to their
respective force matrices.  While the matrix {for second derivative of
  spin with respect to atomic displacement, i.e., the spin force
  matrix, } has a similar structure as the vibrational force matrix,
its diagonal elements are smaller than its off-diagonal elements, so
the spin {dynamical} matrix has both negative and positive
eigenvalues. The spin-phonon dispersion proposed here is expected to
be applicable across all magnetic materials; it points out regions
where the spin properties can be tailored through the phonon
excitation.

The rest of the paper is arranged as follows. In Sec. II, we present
the theoretical formalism, where {\clr we first define the spin
  {dynamical} matrix and then present the details of the spin-phonon
  dispersion calculation.}  Section III provides our results, where
the spin change in hcp Co for two phonon modes is first examined,
followed by the spin-phonon dispersions of hcp Co and bcc Fe. We
explain these main differences using the vibrational force and spin
force matrices.  {\clr Section IV is devoted to the discussion of our
  results and their relation to the ongoing experimental research and
  the Elliot-Yafet theory.} We conclude this paper in Sec. V.

\newcommand{\br}{{\bf r}}

\newcommand{\ik}{i{\bf k}}

\newcommand{\jk}{j{\bf k}}

\newcommand{\lk}{l{\bf k}}

\newcommand{\bk}{{\bf k}}

\section{Theoretical formalism}

\newcommand{\iik}{i,i,{\bf k}}

{\clr Even in nonmagnetic metals, the traditional Elliot-Yafet (EY)
  theory \cite{elliot1954,yafet1963} has shown that phonons play a
  significant role in spin relaxation.  Spin hot spots were identified
  in polyvalent metals \cite{fabian1998,fabian1999}.  In magnetic
  materials, states are spin-polarized and with the presence of
  spin-orbit coupling, one has to introduce the $SD$ factor to
  characterize the spin hot spots in laser-induced demagnetization
  \cite{aip12}, which highlights the fact that a single spin-phonon
  coupling parameter is not enough. }

Here we employ the density functional theory as implemented in the Wien2k
code \cite{wien2k} and VASP code \cite{vasp}, with the details
presented elsewhere \cite{np09,prl12,gu2013,jpcm16,jpcm17c}. In brief, we solve the
Kohn-Sham equation self-consistently \cite{jpcm16}, \be \left
[-\frac{\hbar^2\nabla^2}{2m_e}+v_{eff}(\br) \right ]
\psi_{\ik}(\br)=E_{\ik} \psi_{\ik} (\br),
\label{ks}
\ee where $ \psi_{\ik}(\br)$ and $E_{\ik}$ are the eigenstate and
eigenenergy of band $i$ and ${\bf k}$ point, respectively.  $v_{eff}$
is determined by \be v_{eff}({\bf r})=v({\bf r})+\int \frac{n({\bf
    r}')}{|{\bf r}-{\bf r}'|} d{\bf r}'+v_{xc}({\bf r}), \ee where
$v_{xc}({\bf r})$ is the exchange-correlation potential, $v_{xc}({\bf
  r})=\delta E_{xc}[n]/\delta n({\bf r})$. The spin-orbit coupling is
included through the second variational principle \cite{wien2k}.  We
use the generalized gradient approximation for the
exchange-correlation energy functional. The density $n({\bf r})$ is
computed from $n({\bf r})=\sum_{\ik}\rho_{\iik}|\psi_{\ik}({\bf
  r})|^2$, where $\rho_{\iik}$ is the electron occupation.

The lattice dynamics is governed by the potential energy change around
the equilibrium position of lattice points.  The expansion of the
potential energy $\Phi$ is 
\begin{widetext}
\be \Phi({\bf R})=\Phi_0+\sum_{i,\alpha}
\left( \frac{\partial \Phi}{\partial { R}_{i,\alpha}}\right )_0 \Delta
     { R}_{i,\alpha}+\frac{1}{2}\sum_{i,\alpha;j,\beta}\left (
     \frac{\partial^2\Phi} {\partial { R}_{i,\alpha} \partial {
         R}_{j,\beta}}\right )_0 \Delta { R}_{i,\alpha} \Delta{
       R}_{j,\beta} +\cdots \ee 
\end{widetext}
where ${ R}_{i,\alpha}$ denotes the
     position of atom $i$ along the $\alpha$ direction.  At
     equilibrium, the second term is zero because the {net forces on
       all atoms} should be zero. So the lowest order in the potential
     is the second order (the third term in the equation). The
     second-order potential derivative matrix ${\cal P}$, the
     vibration force matrix, is defined as \be {\cal
       P}_{ij}^{\alpha\beta}=\left (\frac{\partial^2\Phi({\bf
         L}_i-{\bf L}_j)} {\partial { R}_{i,\alpha} \partial {
         R}_{j,\beta}}\right )_0, \ee which is used to compute the
     phonon spectrum.  Here ${\bf L}_i$ and ${\bf L}_j$ are the
     lattice vectors for atoms $i$ and $j$, respectively.  Terms
     higher than the second order are ignored.

\begin{figure}
  \includegraphics[angle=0,width=0.6\columnwidth]{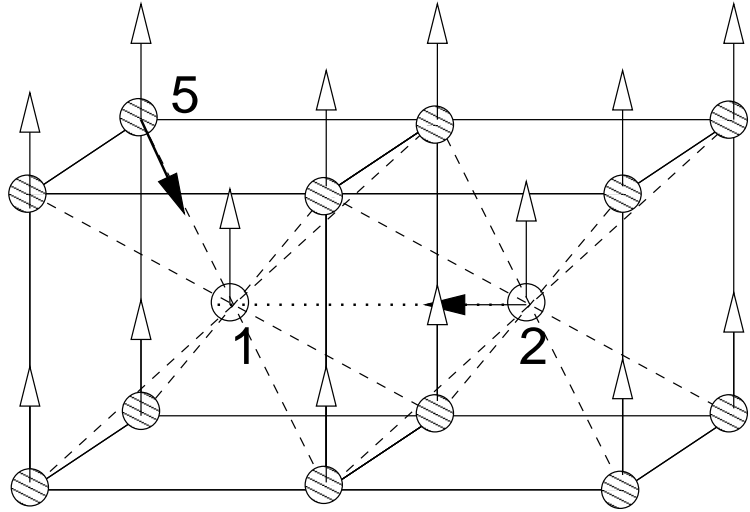}
\caption{An example of the two-cell bcc structure to explain how we
  compute the spin-phonon dispersion. Each atom's spin (empty arrows)
  is initialized along the $z$ axis.  The reference atom 1 is at the
  body center, with the eight first neighboring atoms denoted (shaded
  spheres), where atom 5 is just an example. The reference atom has
  six second-neighbor atoms, where atom 2 is just an example.  For both
  bcc Fe and hcp Co, we adopt a $2\times 2\times 2$ supercell. The
  figure here is just a part of it, without an overly cumbersome
  diagram.  Both the vibrational force and spin force matrices are
  computed by displacing one atom at one time.  Two filled arrows
  denote two unique displacement directions.  }
\label{fig0}
\label{fig1}
\end{figure}

Similar to the lattice vibration, the total magnetic spin moment of a
crystal can be expanded as a function of the displacements of the
atoms, i.e.,
\begin{widetext}
\begin{equation}
M_s
=M_0+\displaystyle\sum_{i\alpha}\left (\frac{\partial
M_s}{\partial R_{i,\alpha}}\right )_0\Delta
R_{i,\alpha}
+\frac{1}{2}\displaystyle\sum_{ij,\alpha\beta}
\left ( \frac{\partial^2 M_s}{\partial R_{i,\alpha}\partial
  R_{j,\beta}}\right )_0 \Delta
R_{i,\alpha} \Delta R_{j,\beta}+... ,
\label{spin}
\end{equation}
\end{widetext}
where $M_s$ is the magnetic moment of the crystal, $M_0$ is the
magnetic moment at the lattice equilibrium, ${\bf R}$ is the position
of the atom, $i(j)$ is the atomic index, and $\alpha (\beta)$ is the
direction index, i.e., $\alpha = x,$ $y, \text{ or } z$.  Different
from the phonon calculation above, the linear term (the second term) $
\left (\frac{\partial M_s}{\partial R_{i,\alpha}}\right )_0 $ is not
zero, because the energy minimum is not the spin moment minimum in a
magnetic material.  The second-order derivative of the spin moment
${\cal S}$, spin force matrix, is defined similarly as, \be {\cal
  S}_{ij}^{\alpha\beta}({\bf L}_i-{\bf L}_j)= \left ( \frac{\partial^2
  M_s({\bf L}_i-{\bf L}_j)} {\partial R_{i,\alpha}\partial
  R_{j,\beta}}\right )_0, \ee which only depends on the relative
lattice vector {\bf L}.  In this work, we use the finite difference
method in a $(2\times2\times2)$ supercell to find the second order
spin moment change. Figure \ref{fig0} shows a two-cell structure for
bcc Fe, where the first nearest neighbor and the second nearest
neighbor atoms are denoted by the dashed and dotted lines,
respectively.  The spin {dynamical} matrix is \be {\cal
  C}^{\alpha\beta}_{ij}({\bf k}) = \sum_{\bf L} {\cal
  S}_{ij}^{\alpha\beta}({\bf L}) e^{-i{\bf k}\cdot {\bf L}},
\label{eq7}
\ee where ${\bf k}$ is the reciprocal lattice vector.  We diagonalize
it to get the spin-phonon dispersion spectrum. The first order change
in spin moment, which is already taken into account in the EY theory
\cite{elliot1954,yafet1963}, only contributes a shift in the
displacement, so it is not included under our current theory.

\section{Results}

Despite the apparent importance of spin-lattice interaction in many
materials \cite{decker2021,andres2021}, except a short presentation
\cite{aps2013}, there has been no prior study on how the spin moment
changes with the lattice. Our study is also different from prior
studies on magnon-phonon coupling \cite{weber1968} where the spin
wave, i.e.  the spatial orientation of the spins, at each lattice site
is coupled to the vibration but the spin amplitude remains the
same. Here, we consider the spin amplitude change when the atoms
vibrate. For this reason, it is necessary to develop a simple picture
before the full scale of calculation.

\subsection{Effects of breathing and stretching modes on the spin
  moment in hcp Co}

 We strategically choose hcp Co as our first example,
because without changing the unit cell size we can already
investigate two vibrational modes: one is the breathing mode, where
the volume is changed while keeping the ratio $c/a$ fixed, and the
other is the stretching mode, where the ratio $c/a$ is changed by
keeping $V$ fixed.  Our in-plane lattice constant $a$ is 4.713924
bohr, and out-of-the-plane one $c$ is 7.651595 bohr.  The unit cell
volume is $V=\sqrt{3}a^2c/2$ \cite{jpcc21}. Here we use the Wien2k
code to carry out our study.  We choose the Muffin-tin radius $R_{\rm
  MT}$ of 2.26 bohr, and its product with the planewave cutoff $K_{\rm
  cutoff}$ is 9. The $k$ mesh is $33\times 33\times 17$, which is
sufficient to converge our results as tested before \cite{jpcc21}.

\begin{figure}
  \includegraphics[angle=0,width=0.7\columnwidth]{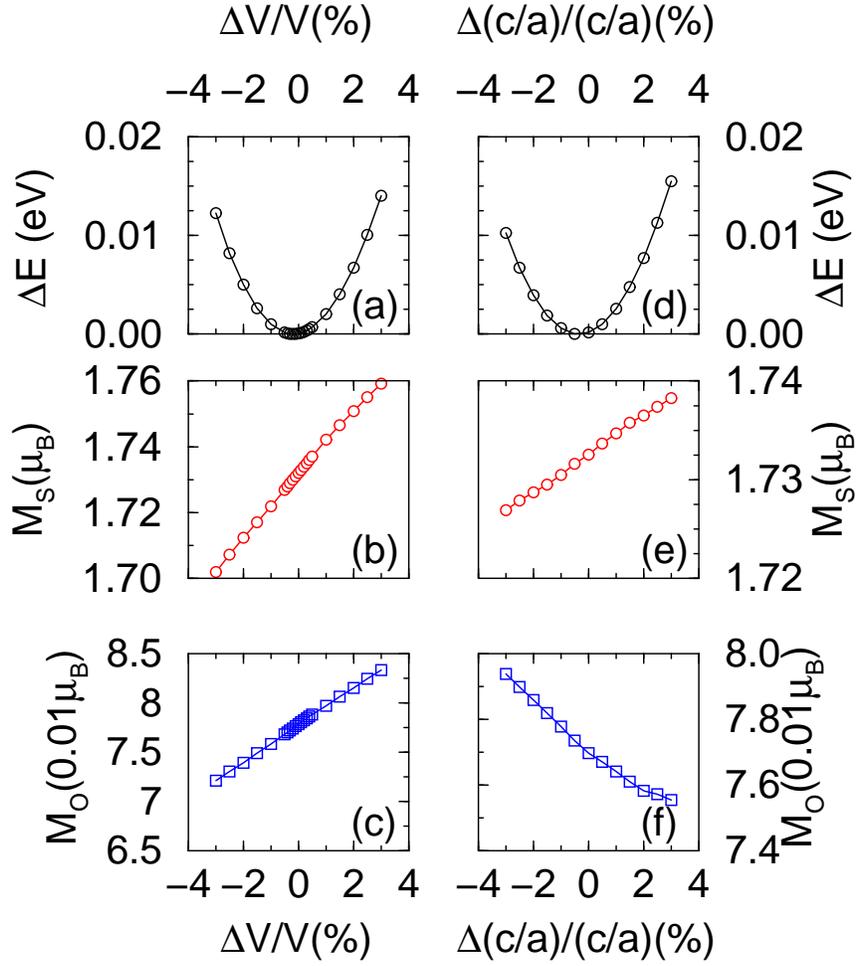}
  \caption{(a) Total energy change $\Delta E$ as a function of $\Delta
    V/V$ for the breathing mode.  $\Delta E$ is referenced with
    respect to the lowest total energy.  The {\bf k} mesh is $33\times
    33 \times 17$.  (b) and (c) are the spin and orbital moment
    changes as a function of $\Delta V/V$, respectively.  (d) Total
    energy change $\Delta E$ for the stretching mode. The {\bf k} mesh
    is much larger, $53\times 53 \times 28$.  (e) and (f) are the spin
    and orbital moment changes as a function of $\Delta(c/a)/(c/a)$,
    respectively.  }
\label{fig2}
  \end{figure}

Figure \ref{fig2}(a) shows the total energy change $\Delta E=E-E_{\rm
  min}$ as a function of volume change $\Delta V/V$ for the breathing
mode, where $E_{\rm min}$ is the energy minimum.  We see that the
entire change is very smooth and forms a typical parabola with a
minimum around -0.2\%, which shows that our starting lattice constant
is slightly too large, but this does not affect our conclusion.  By
contrast, the spin moment ($M_S$) change increases linearly with
$\Delta V/V$ (see Fig.  \ref{fig2}(b)).  For every 1\% change in
$\Delta V/V$, the spin changes by about 0.01 $\mu_B$, which can be
fitted to $M_{S}=\left [ 1.73+0.00965 \Delta V/V-0.0001469 (\Delta
  V/V)^2 \right ] ~\mu_B$, where the second term is the linear
contribution.  Figure \ref{fig2}(c) shows the orbital moment change
with $\Delta V/V$, which can be fitted to $M_{O}=(0.0778+
0.00189\Delta V/V)\mu_B$.  The orbital change is only 1/5 the spin
change.  The results from the stretching mode are shown in
Figs. \ref{fig2}(d), (e), and (f).  Figure \ref{fig2}(d) shows that
the stretching mode has a similar smooth energy change as a function
of $c/a$.  Its spin change (Fig. \ref{fig2}(e)) is smaller than that
in the volume change; and for one percentile change in $c/a$, the spin
changes by 0.001895 $\mu_B$. The second order coefficient is below
$10^{-6}$.  The orbital $M_O$ behaves very differently from the spin:
it decreases with $c/a$ (Fig.  \ref{fig2}(f)), and the change is not
strictly linear.  So far the spin changes are limited to two lattice
modulations.  To go beyond this, we need to introduce the spin-phonon
dispersion.

\subsection{Spin-phonon dispersion}

\begin{figure}
\includegraphics[angle=0,width=0.7\columnwidth]{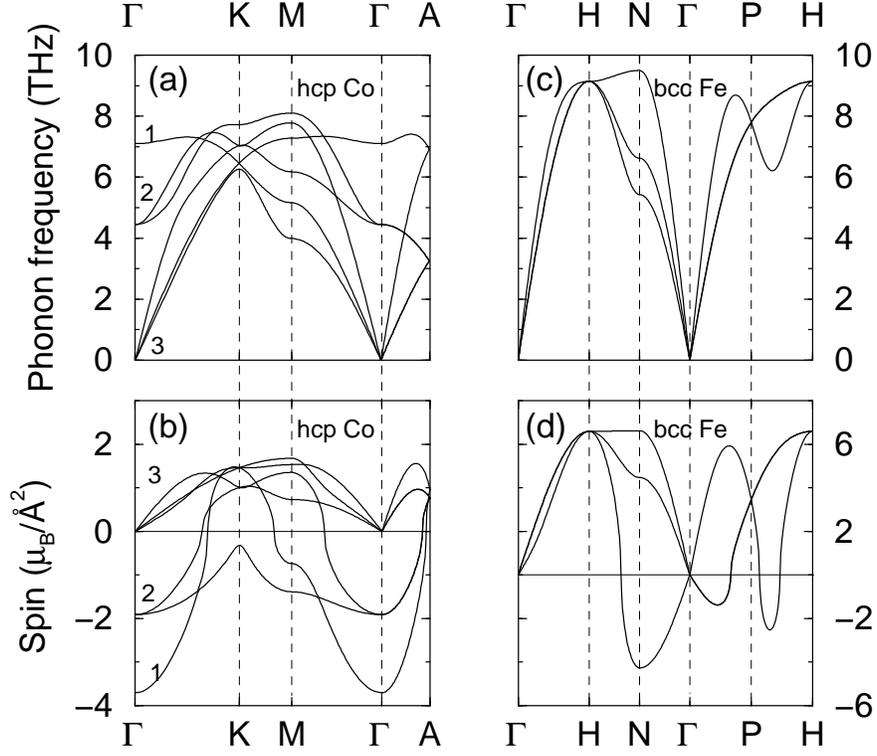}
\caption{ (a) Phonon spectrum of hcp Co.  (b) Spin-phonon dispersion
  for hcp Co.  The total spin moment is expanded up to the second
  order. Both the positive and negative spin moment changes are
  noticed. Numbers around the $\Gamma$ point in (a) and (b) denote the
  degeneracy of the bands.  (c) Phonon spectrum of bcc Fe. (d)
  Spin-phonon dispersion for bcc Fe.  }
\label{fig3}
\label{co}
\label{fe}
\end{figure}

We start with hcp-Co. There have been many studies on the phonon
dispersion of hcp Co. For instance, Pun and Mishin \et \cite{pun2012}
employed the embedded-atom potential to compute the phonon spectrum.
We employ the VASP code \cite{vasp} and the frozen-phonon method and
our computed phonon dispersion is shown in Fig. \ref{co}(a). The
frequencies of the two optical modes at $\Gamma$ point are 4.45
(doublet) and 7.12 (singlet) THz, respectively, which match the
neutron scattering results \cite{wakabayashi1982}  of 4.30 and 7.60
THz, respectively.  This agreement gives us confidence that our theory
can give an accurate prediction on the lattice dynamics.

We then move on to the spin moment change as a function of lattice
vibration. As predicted by Eq. \ref{spin}, the spin has a nonzero
second-order term in the lattice displacement.  We make the same
Born-Oppenheimer approximation for the spin where the lattice responds
much slower than the electron spin, an assumption that needs
experimental verification.  Then we can apply the same frozen-phonon
method to the spin moment change as a function of atomic displacement.
By diagonalizing the spin {dynamical} matrix (Eq. \ref{eq7}), we
obtain the spin dispersion as a function of the crystal momentum.
Physically, this represents the spin moment change as the atoms
vibrate according to the normal modes of phonons.

Figure \ref{co}(b) is our spin dispersion for hcp Co. {The high symmetry
points are highlighted with the dashed vertical lines.} The general
structure follows the phonon (Fig. \ref{co}(a)), but the order of the
spectrum is different. At the $\Gamma$ point, the phonon spectrum has
three acoustic bands at the bottom but the spin has them at the
top. In addition, the degeneracy increases from 1 to 3 from the bottom
to the top. The horizontal line sets the zero value for the spin
dispersion.  This feature persists from $\Gamma$ to K, but at the M
point, similar to the phonon spectrum, no degeneracy is found. At the
A point, all the bands are close to a single point.  

The major difference from the phonon spectrum is that the spin
dispersion has several branches with “imaginary modes” (negative
eigenvalues).  Just as the imaginary phonon modes leading to the
structural instability, the spin imaginary modes lead to spin
reduction or demagnetization. The activation of those modes reduce the
magnitude of the total spin moment.  This phonon-induced
demagnetization process is different from the conventional magnon
picture or thermal-based demagnetization mechanism where the spin
spatial orientation change reduces the spin moment for the entire
sample.  In our case, the magnitude of the local spin moment is
changed as the atom vibrates along the eigenvector of a phonon mode.
Quantitatively, if we examine Fig. \ref{co}(b) closely, we notice that
the LO mode at the $\Gamma$ point has a larger contribution to the
reduction of spin moment than that of the two fold degenerate TO
modes. The reason why these modes behave differently can be explained
by examining the Co-Co bond. Physically, the longitudinal mode
shortens the bond length on one side, while elongates it on the
other. When the distance between neighboring atoms decreases, the
increased electron overlap broadens the 3d bandwidth and reduces the
density of state at the Fermi level. This mechanism is also the origin
of the suppression of the magnetization when the lattice constant of
Co reduces. By contrast, the transverse mode acts differently. Instead
of changing the bond length, it distorts the angles between lattice
vectors, i.e. changing the bond angles.  In other parts of the
Brillouin zone, both the spin increase and decrease are found.  This
is our first major finding.

This spin moment change is not limited to hcp Co.  In bcc Fe, we find
a similar pattern. Figure \ref{fe}(c) is our phonon dispersion. With
fewer atoms in the cell, the number of bands is reduced.  Our
spin-phonon dispersion is shown in Fig. \ref{fe}(d).  Between the
$\Gamma$ and H points, the spin-phonon and the phonon dispersions are
almost identical, but the major difference appears around the N point,
where we see a major negative eigenvalue branch. This corresponds to
the spin moment reduction.  However, the spin enhancement dominates
along these high symmetry lines. The $k$-point convergence test has
been performed up to $(32\times32\times32)$. The data in the paper
uses the $k$-point mesh of $(16\times16\times16)$.


\subsection{Insights into the spin-phonon dispersion}

To have a deeper understanding of the phonon-induced spin moment
change, we compare the vibrational force matrix ${\cal P}_{ij}^{\alpha
  \beta}$ and the spin force matrix ${\cal S}_{ij}^{\alpha \beta}$,
where $i$ and $j$ are the atom indices and $\alpha \beta$ are the
Cartesian coordinate indices.  These two matrices are used to compute
the phonon and spin-phonon dispersions.  We take bcc Fe as an
example. For the supercell $2\times 2\times 2$, eight atoms are at
positions, $(0,0,0), (\frac{1}{2},\frac{1}{2},0),
(\frac{1}{2},0,\frac{1}{2}), (0,\frac{1}{2},\frac{1}{2}),
(\frac{1}{2},0,0),$ $ (0,\frac{1}{2},0), (0,0,\frac{1}{2}),
(\frac{1}{2},\frac{1}{2},\frac{1}{2})$.  So there are 64 combinations,
leading to sixty-four $3\times 3$ matrices along the $x$, $y$ and $z$
axes. The simplest matrix is the on-site one.
\begin{equation}
{\cal S}_{1,1}=\left(
\begin{array}{rrr}
    \gamma &     0 &     0 \\
      0 &   \gamma &     0 \\
      0 &     0 &    \gamma \\
\end{array}
\right );\hspace{1cm}
{\cal P}_{1,1}=\left(
\begin{array}{rrr}
g & 0 & 0 \\
0 & g & 0 \\
0 & 0 & g \\
\end{array}
\right ).
\end{equation}

It is clear that the spin and vibrational force matrices have the same
structure. This is true for four matrices for the nearest neighbors in
the same primitive cell, 
\begin{equation}
{\cal S}_{1,5}=\left(
\begin{array}{rrr}
   -\alpha & \beta &\beta  \\
  \beta &  -\alpha &-\beta \\
  \beta &-\beta &   -\alpha \\
\end{array}
\right );\hspace{1cm}
{\cal P}_{1,5}=\left(
\begin{array}{rrr}
   -a & b &b  \\
  b &  -a &-b \\
  b&-b &   -a \\
\end{array}
\right ),
\end{equation}
but we notice a crucial difference. In the spin force matrix, the
diagonal element $\alpha$ is always smaller than the off-diagonal
element $\beta$, where $\alpha$ and $\beta$ are matrix elements, not
to be confused with the direction indices above.  Numerically, if we
use those ${\cal S}_{1,5}$ matrices alone, we end up to have a
negative eigenvalue.  This explains why the spin-phonon dispersion has
a negative eigenvalue, but the phonon one does not. 
The other three matrices are found by cyclically exchanging the
off-diagonal elements.

Another major difference is in the 
 matrices for the next nearest neighbor cell. 
\begin{equation}
{\cal S}_{1,2}=\left(
\begin{array}{rrr}
    \delta &     0 &     0 \\
      0 &   \delta &     0 \\
      0 &     0 &    \epsilon \\
\end{array}
\right );\hspace{1cm}
{\cal P}_{1,2}=\left(
\begin{array}{rrr}
-d & 0 & 0 \\
0 & -d & 0 \\
0 & 0 & -e \\
\end{array}
\right ).
\end{equation}
The other two matrices can be reproduced by the cyclic relation. One
sees that the spin force matrix has a positive value, but the
vibrational force matrix is all negative. Since both the vibrational
force matrix and the spin force matrix must obey the sum rule, where
the summation over the cells and atoms is zero, we find that for the
spin, $\gamma+2\delta +1\epsilon-4\alpha=0$, but for the vibration,
$g-2d-1e-4a=0$.  {\clr All the parameters are given in Table
  \ref{tab1}.}

\section{Discussions and implication for experiments}

\begin{figure}
\includegraphics[angle=0,width=0.7\columnwidth]{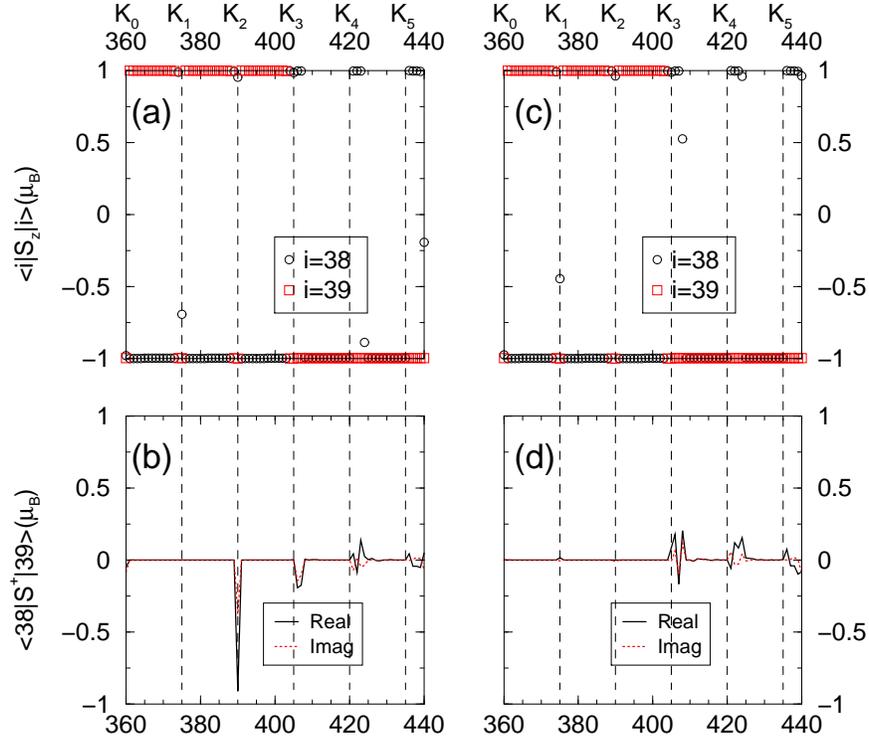}
\caption{\clr (a) Spin expectation values of states $i=38$ (circles)
  and $i=39$ (boxes) along the crystal momentum for five segments with
  $k_1$ through $k_5$, whose coordinates are given in the main
  text. The vertical dashed lines denote their locations.  The numbers
  on the $x$ axis are the $k$ index of the $k$ list, with the $k$ mesh
  $(53\times 53\times 28)$.  (b) Spin flip matrix element $s^+$ $\la
  38|s^+|39\ra$ between states 38 and 39. The element for $s^-$ is
  tiny, not shown.  The solid and dashed lines denote the real and
  imaginary parts, respectively.  (c) and (d) are the same as (a) and
  (b), but for the distorted structure, with $-0.5\%$ contraction as
  Figs. \ref{fig2}(d), (e) and (f). This mimics the temperature
  effect.  (c) shows the spin change is concentrated in some hot spin
  spots.  (d) shows a larger spin flip change due to the lattice
  distortion.  Even the location of the larger spin flip is
  changed. The largest change is now at $k_3$.  }
\label{fig4}
\end{figure}

{\clr In nonmagnetic metals, the spin lifetime $\tau_s$ is equal to
  the electron momentum relaxation time multiplied by the EY constant
  $\beta$, which is the ratio of the energy gap $\Delta E$ to the
  spin-orbit coupling $\lambda_{soc}$ \cite{elliot1954,yafet1963}: \be
  \tau_s=\beta \tau_e =\left (\frac{\Delta E}{\lambda_{soc}}\right )^2
  \tau_e \ee where the energy gap $\Delta E$ is between two scattering
  states. For multiple channels of scattering, the inverse spin
  lifetime $1/\tau_s$, or the rate of spin relaxation, is computed
  from \be \frac{1}{\tau_s} =\sum_i \frac{1}{\beta_i\tau_{e,i}}.  \ee
  This relation is controlled by the smallest $\beta_i$, which is
  often the phonon, $\beta_{phonon}$. Fabian and Das Sarma
  \cite{fabian1998} further showed in polyvalent metals, such as Al,
  that the spin relaxation is significantly increased around spin hot
  spots \cite{fabian1999}, where a Fermi surface cuts through the
  Brillouin zone boundaries and special symmetry points and
  lines. Experiments showed that $\beta$ is not even a constant and
  depends on surfaces and interfaces \cite{watts2022}.  In magnetic
  materials, this is much more complicated \cite{aip12} and also more
  important for spintronics.

Here, we make a moderate attempt to show that the lattice distortion
affects the spin and spin flip among the band states, with a detailed
study left for the future due to the enormous complications in
magnetic materials. We choose two states close to the Fermi level,
band states 38 and 39. Due to the band dispersion, we cannot guarantee
that they always stay close to the Fermi level.  All the conclusions
drawn here are specific to these two states, and should not be applied
to others without a detailed calculation.  Initially, we aim to choose
a special symmetry line, from $\Gamma$ to A, along the $k_z$ axis, but
it turns out that due to the momentum locking in the spin-orbit
coupling, along this direction there is almost no spin flip. Since
showing results for a $k$ mesh of $(53\times 53\times 28)$ is overly
excessive, we decide to choose five segments which are not along any
high symmetry lines.  The $k$ only changes along the $z$ axis from 0
to 0.5 in the units of the reciprocal lattice vector, so we can
monitor the pattern of change. Other than this, the $k$ choice is
arbitrary.  Segment 1, from 361 to 375, has the $k$ change from
$k_0=(0,0.45,0)$ to $k_1=(0,0.45,0.5)$, where the numbers 361 to 375,
as well as those on the $x$ axis of Fig. \ref{fig4}, are the $k$ index
of the $k$ list out of $(53\times 53\times 28)$.  Segment 2, from 376
to 390, has the $k$ change from $(0,0.4716,0)$ to
$k_2=(0,0.4716,0.5)$; segment 3, from 391 to 405, has the $k$ change
from $(0,0.4905,0)$ to $k_3=(0,0.4905,0.5)$; segment 4, from 406 to
420, has the $k$ change from $(0,0.01886,0)$ to $k_4=(0,0.01886,0.5)$;
and segment 5, from 421 to 435, has the $k$ change from
$(0.001886,0.0377,0)$ to $k_5=(0.001886,0.0377,0.5)$.

Figures \ref{fig4}(a) and (b) show the spin moments $\la nk|s_z|nk\ra$
of states 38 and 39 and the spin flipping $\la nk|s^+|mk\ra$ and $\la
nk|s^-|mk\ra$ between them, where $s_z$ is the $z$-component of the
spin operator, and $s^\pm$ are the raising and lowering spin
operators, respectively.  Figure \ref{fig4}(a) reveals that most of
the $k$ points have spin moment close to either $+1\mu_B$ or
$-1\mu_B$. These spin moments are not exactly at $+1\mu_B$ or
$-1\mu_B$; with the spin-orbit coupling, either $+1\mu_B$ or $-1\mu_B$
is not allowed for states with a nonzero orbital angular momentum. In
the first segment between $k_0$ and $k_1$, state 38 has spin moment
close to $-1$ $ \mu_B$, and state 39 has spin moment close to $+1$ $\mu_B$,
but this is not enough to have a spin flip.  Figure \ref{fig4}(b)
shows that the spin flip between them is small because the spin flip,
the expectation value of $\la 38|s^+|39\ra$, critically depends on the
product of the small spin component of one state's wavefunction with
the large spin component of another state.  $\la 38 |s^-|39\ra$ is not
shown since it is very small for this pair of states.  The first
maximum occurs at the end of the first segment $k_2$, where the spin
of state 28 drops to $-0.69$ $\mu_B$. A smaller spin flip is at $k_3$
and $k_4$. If we slightly distort the lattice in the same fashion as
Figs. \ref{fig2}(d), (e) and (f) (the lattice change is -5\%), we can
investigate the effect of the lattice vibration. Figure \ref{fig4}(c)
shows that while the majority of the spins remain similar to
Fig. \ref{fig4}(a), the hot spin pockets \cite{aip12} are relocated to
different $k$ points. A bigger change is in the spin flip, with the
maximum shifted to $k_3$. This shows that the spin flipping is very
sensitive to the lattice perturbation.  Numerically we find that the
initial wavefunctions that form the basis of the spin-orbit coupling
in the second-variational step have a significant impact on the phase
of the matrix elements of $s^+$ and $s^-$, but the effect on the $s_z$
is tiny.  Our results may change if the wavefunction is converged at a
different criterion. In our calculation, we use an extremely low
charge convergence of 10$^{-7}$.  We should point out that
Fig. \ref{fig4} only samples a small portion of the Brillouin zone and
one should not conclude that the distorted structure has a smaller
spin flip, since a significant variation of the spin flip in
comparison to the undistorted structure is found across the entire
Brillouin zone. Our goal here is to show that the phonons do affect
the spin flip, highlighting the importance of the spin-phonon
dispersion.}

The negative eigenvalues found in the spin-phonon dispersion identify
a demagnetization channel through the nearest neighbor vibration,
which could have some implications on the recent debate on the
mechanism of femtomagnetism \cite{eric,ourreview,rasingreview}. It has
been argued that the phonon plays a significant role in
demagnetization \cite{dornes2019,tauchert2022}. While we do not
consider the electronic excited state, the spin-phonon dispersion
demonstrates unambiguously that not every phonon excitation leads to
demagnetization. If we assume a harmonic oscillation of lattice with
time as shown by Henighan \et \cite{henighan2016}, we expect that the
spin oscillates as $M_{S}(t)=M_{S}^{(0)}+M^{(1)}_{S}\sin(\Omega t)$,
where $t$ is the time, $M_{S}^{(0)}$ and $M_{S}^{(1)}$ are two
constants, and $\Omega$ is the phonon frequency. This does not
necessarily correspond to demagnetization. In hcp Co, Fig. \ref{co}(b)
shows that the maximum negative eigenvalue reaches $-3.7$
$\mu_B/\rm\AA^2$ at $\Gamma$, which corresponds to the optical phonon
mode in Fig. \ref{co}(a).  If we suppose the phonon mode displacement
to be 0.01 $\rm \AA$, which is typical in solids, this leads to the
spin moment reduction of $-3.7 \times 10^{-4}$ $ \mu_B$, which is
quite small. However, there are at least two cases that the spin
reduction can be boosted. {\clr One is the multiphonon generation
  which is featured by the THz anharmonic frequency shift
  \cite{feng2018,takahashi2022}. Under the same temperature,
  lower-frequency phonon modes are generated more.  At an elevated
  temperature, we expect that the lattice vibration anharmonicity
  appears since the coupling between the phonon and demagnetization
  becomes highly nonlinear.  This may contribute additional amounts of
  demagnetization.}  The second case is if one employs a strong laser
pulse, where the displacement can reach 0.5 $\rm \AA$. In bcc Fe, even
though only one pocket has a larger negative eigenvalue (around $N$),
its value, $-4.4$ $\mu_B\rm/\AA^2$, is more negative than that in hcp
Co. This suggests a scenario for future experiments. If one can
selectively excite a special phonon mode, through Raman or neutron
scattering \cite{mook1966}, and then detect the spin moment change,
one may be able to completely map out the entire spin-phonon
dispersion.

{\clr Finally, we note that the phonon coupling to other degrees of freedom
has been observed in nanomagnets \cite{berk2019}, Gd \cite{sultan2012}
and other magnetic systems
\cite{cheng2008,mikhail2011,verstraete2013,li2021,raut2022}.  Shin \et
\cite{shin2021} showed even a quasi-static strain plays an important
role in ultrafast spin dynamics.  Rongione \et \cite{rongione2022}
detected a bipolar strain wave-induced THz torque in the (001)
oriented NiO/Pt film where the stress from the Pt layer launches the
strain wave.  Recently, Mashkovich \et \cite{mashkovich2021} employed
the strong spin-lattice coupling in antiferromagnetic CoF$_2$ to
excite phonons via magnons. For the same CoF$_2$ system, Disa \et
\cite{disa2020} did the opposite. They excited two degenerate $E_u$ IR
modes to displace the lattice along the optically inaccessible
$B_{2g}$ Raman mode, which transiently breaks the site symmetry
between two Co atoms. The spin moments on two Co atoms do not
compensate each other, and CoF$_2$ temporally transitions to a
ferrimagnet, with a net spin moment of 0.21 $\mu_B$.  The spin-phonon
dispersion introduced here is expected to capture many opportunities
in the entire Brillouin zone. This points out a new direction -- the
spin-phonon dispersion in excited states. Excited states often
accommodate strong THz anharmonicity, leading to anharmonic frequency
shifts through multiple phonon scattering.}

\section{Conclusions}

We have introduced a much needed concept of the spin-phonon
dispersion, where the second-derivative of the spin moment change is
dispersed along the phonon crystal momentum. Our spin-phonon
dispersion, with spin and phonon information both present, captures
the complexity of the dependence of spin on the phonon, which goes
beyond the spin-phonon coupling constant.  The spin-phonon dispersion
in hcp Co shows that a large portion of the Brillouin zone
accommodates both spin reduction and increase, but this is not the
case for bcc Fe. We only find a small pocket around the N point in bcc
Fe that has spin moment reduction. This may help us understand the
higher Curie temperature in Fe. Microscopically, we also understand
why the spin-phonon dispersion has negative eigenvalues.  In contrast
to the vibrational force matrix, the off-diagonal matrix elements in
the spin force matrix are larger than its diagonal counterparts, so
the spin {dynamical} matrix contains the negative eigenvalues.  Our
spin-phonon dispersion is also different from the magnon-phonon
coupling \cite{weber1968} where the spin orientation, not the spin
magnitude, is taken into account.  {Finally, our approach can be
  easily extended to systems where spin-orbit coupling should be taken
  into account}. Thus, the spin-phonon dispersion represents an
important paradigm shift and will have a significant impact on future
research in both simple and complex quantum magnetic materials in
spatial and time domains \cite{eric}.

\acknowledgments 

We acknowledge the helpful comments on our paper from Dr.  Qihang Liu
(SUST, China).  MG was supported by Foundation for Distinguished Young
Talents in Higher Education of Guangdong under Grant No. 2020KQNCX064,
Shenzhen Science and Technology Innovation Council under Grant
No. JCYJ20210324104812034, and Natural Science Foundation of Guangdong
Province under Grant No. 2021A1515110389.  GPZ and YHB were supported
by the U.S. Department of Energy under Contract No.
DE-FG02-06ER46304.  Part of the work was done on Indiana State
University's quantum and obsidian clusters.  The research used
resources of the National Energy Research Scientific Computing Center,
which is supported by the Office of Science of the U.S. Department of
Energy under Contract No. DE-AC02-05CH11231.

$^*$guo-ping.zhang@outlook.com
https://orcid.org/0000-0002-1792-2701

\begin{table}
\caption{\clr Spin ($\alpha,\beta,\gamma,\delta,\epsilon$) and phonon
  ($a,b,g,d,e$) parameters used for bcc Fe to compute the spin-phonon
  and phonon spectra, respectively, with 
  $\alpha,\beta,\gamma,\delta,\epsilon$ in the units of
  $\mu_B/\rm\AA^2$. }
\begin{tabular}{ccccc}
\hline
\hline
$\alpha$ & $\beta$ & $\gamma$ & $\delta$ &$\epsilon$ \\
1.4232 &  2.0323 & 4.4081 &  0.1126 & 1.0594 \\
\hline
$a$ & $b$ &$g$& $d$ &$e$ \\
2.3876 & 1.736 & 11.7342 &0.1163 &1.9512\\
\hline
\hline
\end{tabular}
\label{tab1}
\end{table}

\end{document}